\documentclass[letterpaper, 10 pt, conference]{ieeeconf}
\IEEEoverridecommandlockouts 
\usepackage[left=50pt,right=50pt,top=60pt,bottom=48pt]{geometry}
\usepackage{cite}
\usepackage{amsmath,amssymb,amsfonts}
\usepackage{graphicx}
\usepackage{caption}
\usepackage{floatrow}
\usepackage{subcaption}
\usepackage{textcomp}
\usepackage{amsfonts, amsmath}
\usepackage{amssymb}
\usepackage{cleveref}
\usepackage{amsmath}
\usepackage{graphicx}
\usepackage{multirow}
\usepackage{etoolbox}
\makeatletter
\patchcmd{\@makecaption}
  {\scshape}
  {}
  {}
  {}
\makeatother

\newtheorem{theorem}{Theorem}

\newtheorem{remark}{Remark}
\newtheorem{assumption}{Assumption}

\usepackage{cuted}
\usepackage[english]{babel}
\usepackage{xcolor}
    
    \makeatletter
 \let\old@ps@headings\ps@headings
 \let\old@ps@IEEEtitlepagestyle\ps@IEEEtitlepagestyle
 \def\confheader#1{%
 \def\ps@headings{%
 \old@ps@headings%
 \def\@oddhead{\strut\hfill#1\hfill\strut}%
 \def\@evenhead{\strut\hfill#1\hfill\strut}%
 }%
 \def\ps@IEEEtitlepagestyle{%
 \old@ps@IEEEtitlepagestyle%
 \def\@oddhead{\strut\hfill#1\hfill\strut}%
 \def\@evenhead{\strut\hfill#1\hfill\strut}%
 }%
 \ps@headings%
 }
 \makeatother

 \usepackage[pscoord]{eso-pic}

  \makeatletter
\newcommand{\linebreakand}{%
  \end{@IEEEauthorhalign}
  \hfill\mbox{}\par
  \mbox{}\hfill\begin{@IEEEauthorhalign}
}
\makeatother  
  
\begin{document}

\title{State Constrained Model Reference Adaptive Control with Input Amplitude and Rate Limits
}

\author{Poulomee~Ghosh and Shubhendu~Bhasin
\thanks{Poulomee Ghosh and Shubhendu Bhasin are with the Department of Electrical Engineering, Indian Institute of Technology Delhi, New Delhi, India. 
        {\tt\small (Email: Poulomee.Ghosh@ee.iitd.ac.in, sbhasin@ee.iitd.ac.in)}}}




\maketitle

\begin{abstract}
This paper proposes a robust model reference adaptive controller (MRAC) for uncertain multi-input multi-output (MIMO) linear time-invariant (LTI) plants with user-defined constraints on the plant states, input amplitude, and input rate. The proposed two-layer barrier Lyapunov function (BLF)-based control design considers the input and the input rate as states that are constrained using two BLFs in the first layer, while another BLF in the second layer constrains the plant states. The adaptive control law ensures that the plant states, input amplitude, and input rate remain within the user-defined safe sets despite unmatched bounded disturbances. Sufficient conditions for the existence of a feasible control policy are also provided. To the best of the authors' knowledge, this is the first optimization-free method that imposes user-defined constraints on the state, input, and input rate and also provides verifiable feasibility conditions in the presence of parametric uncertainties and disturbances. Simulation results demonstrate the effectiveness of the proposed algorithm.


\end{abstract}


\section{Introduction}
\label{sec:intro}
Safety-critical control systems have risen in prominence in recent years, with safety now considered a major design goal, sometimes even more important than stability. The goal of system safety can often be translated to constraints on the states and inputs. The control problem is further complicated in the presence of system uncertainty. Model Reference Adaptive Control (MRAC), a popular adaptive control approach for systems with parametric uncertainty \cite{classMRAC}, is typically not designed to adhere to user-defined state and input constraints, which often limits its practical applicability.
Therefore, the problem of designing an MRAC algorithm for uncertain plants that ensures tracking to a desired reference model while constraining the state, input magnitude, and input rate is both practically relevant and theoretically challenging.\\
Several control strategies have been proposed to tackle these issues, either partially or completely. Output-constrained adaptive tracking control has been addressed through techniques such as model predictive control (MPC) \cite{mpcnew}, optimal control\cite{opt}, invariant set theory \cite{blanchini}, reference governor approach \cite{rga}, control barrier function (CBF)-based approach \cite{CBF} etc. However, most of these approaches necessitate running an optimization routine at each time step, which may become computationally demanding and lead to challenges in stability analysis.
To address these concerns, barrier Lyapunov function (BLF) \cite{BLF, uub_blf} is being widely used to ensure safety; however, employing BLF to tackle user-defined constraints on the control input is not yet fully explored in the literature.\\
Existing literature has extensively considered the practical challenge of constraining inputs to accommodate actuator limits, especially by employing various saturated functions, such as hyperbolic tangent, sigmoid, etc. \cite{incon10, inconnew}. 
Furthermore, in \cite{annaswamy}, \cite{Lav}, the plant uncertainty is handled by using an adaptive tracking controller for a single-input, single-output (SISO) linear time-invariant (LTI) plant with input constraints. \\
The above-mentioned approaches typically handle either state or input constraints. Few control strategies exist that effectively handle both state and input constraints for uncertain systems. MPC \cite{mpcnew, dhar} stands out as a popular method that integrates constraints on state and input within the optimization process, albeit with increased computational complexity and conservatism, especially in the presence of model uncertainty. 
Approximation-based adaptive controllers along with a BLF and saturation control are used in \cite{fuzzy1, fuzzy2} to tackle both input and output constraints using an adaptive neural network or fuzzy logic-based approach to deal with nonlinearity. A few recent works \cite{Anderson, chattopadhyay2025model} develop MRAC laws that place user-defined bounds on state and input by either developing an auxiliary reference model or modifying the reference model; however, they do not consider constraints on the input rate.
\\
Additionally, apart from constraints on input amplitude, rate limit has also emerged as a crucial concern in control practice. In numerous scenarios, limitations on the rate of actuator force can induce severe performance deterioration or even instability\cite{zaccarian2011modern}. Several approaches, including convex computational method \cite{kapila1999lmi}, anti-windup strategy \cite{turner2020anti}  have been extensively used to tackle input magnitude and rate saturation.
\\
In our previous work \cite{mymrac, myel}, we designed an adaptive controller to impose user-defined bounds on the plant state and control input magnitude; however, no feasibility conditions were provided, and it was assumed that the
required control effort was always within the input limit. In the proposed approach, we relax this assumption and establish verifiable feasibility conditions for the state, input, and input rate bounds in the presence of bounded external disturbances. Subsequently, an MRAC architecture is designed for an uncertain multivariable LTI system that ensures reference model tracking while also adhering to user-defined constraints on the plant states and the magnitude and rate of the control input. \\
The analysis of systems with input rate constraints is challenging since it typically involves working with the derivative of the state equation. To circumvent this issue, we consider the input and the input rate as states, and a two-layer BLF-based design is used to ensure constraint satisfaction. In the first layer, two BLFs are used to impose different user-defined bounds on the input amplitude and the input rate. The results from this layer are then leveraged in the second layer, where another BLF is methodically integrated to ensure that the plant states remain bounded within a user-defined constraint set. The control law is made robust to external disturbances by using $\sigma$-modification, yielding a uniformly ultimately bounded (UUB) result. Conditions for a constraint-compliant feasible control policy are provided that impose a lower bound on the state constraint. \\
The primary contributions of the work are:
\begin{itemize}
    \item [(i)] The proposed optimization-free adaptive control method addresses a gap in the literature by simultaneously considering state, input magnitude, and input rate constraints for multi-input multi-output (MIMO) LTI systems in the presence of parametric uncertainties and bounded external disturbances.
    \item [(ii)] The two-layer BLF-based approach provides a unified framework for handling multiple constraints (state, input magnitude, and input rate) simultaneously. This integrated approach results in a smooth control law, potentially avoiding any undesired transient behavior.
    \item [(iii)] Sufficient conditions are provided to check the existence of a feasible control policy under user-defined constraints.
\end{itemize} 
 Throughout this paper, \(\mathbb{R}\) represents the set of real numbers, and \(\mathbb{R}^{p \times q}\) refers to the set of \(p \times q\) real matrices and the identity matrix in \(\mathbb{R}^{p \times q}\) is denoted as \(\mathbb{I}_{p\times q}\). The Euclidean vector norm and its induced matrix norm are represented by \(\|\cdot\|\). For a matrix \(A \in \mathbb{R}^{n \times n}\), the trace is denoted by \(\text{tr}(A)\), the largest real part of its eigenvalues by \(\max(\lambda_{\Re}\{A\})\), and the eigenvalues with the smallest and largest real parts by \(\lambda_{\text{min}}\{A\}\) and \(\lambda_{\text{max}}\{A\}\), respectively.


\section{Problem Formulation}
\label{sec:LFSR}


Consider the dynamics of the following MIMO LTI system
\begin{align}
    \dot{x}=Ax+Bu+d
    \label{plant}
\end{align}
Here, \(x(t) \in \mathbb{R}^n\) represents the system state, and \(u(t) \in \mathbb{R}^m\) is the control input, which is continuous and twice differentiable. The matrix \(A \in \mathbb{R}^{n \times n}\) is an unknown system matrix which is assumed to be Hurwitz, while \(B \in \mathbb{R}^{n \times m}\) is a known input matrix, assumed to be full column rank. The system pair \((A, B)\) is assumed to be stabilizable. $d(t)\in \mathbb{R}^n$ represents the unknown unmatched external disturbance which is bounded such that $\|d(t)\|<\bar{d}$, where $\bar{d}>0$ is assumed to be known.\\
A stable reference model dynamics is considered
\begin{align}
    \dot{x}_r=A_rx_r+B_rr
    \label{ref}
\end{align}
Here, \(x_r(t) \in \mathbb{R}^n\) represents the reference model state, and \(r(t) \in \mathbb{R}^{m}\) is a piecewise continuous reference input, uniformly bounded by \(\|r(t)\| < \bar{r}\), where $\bar{r}$ is a known positive constant. The matrices \(A_r \in \mathbb{R}^{n \times n}\) and \(B_r \in \mathbb{R}^{n \times m}\) are known and controllable, with \(A_r\) being Hurwitz, i.e., for any symmetric positive definite matrix \(Q \in \mathbb{R}^{n \times n}\), there exists a symmetric positive definite matrix \(P \in \mathbb{R}^{n \times n}\) satisfying \(A_r^T P + P A_r + Q = 0\).\\ 
 The control objective is to design a feasible control input $u(t)$ that ensures the plant state $x(t)$ tracks the reference model state $x_r(t)$ as $t \rightarrow \infty$  while adhering to the following constraints on plant states, input magnitude, and input rate. \\
 \textbf{State constraint:} The plant states are uniformly bounded, i.e. $\|x(t)\|< \bar{\mathcal{X}}$ $\forall t \geq 0$, where $\bar{\mathcal{X}}$ is a user-defined positive constant.\\
 \textbf{Input amplitude constraint:} The amplitude of the control input is uniformly bounded, i.e. $\|u(t)\|< \bar{\mathcal{U}}_1$ $\forall t \geq 0$, where $\bar{\mathcal{U}}_1$ is a user-defined positive constant.\\
 \textbf{Input rate constraint:} The rate of the control input is uniformly bounded, i.e. $\|\dot{u}(t)\|< \bar{\mathcal{U}}_2$ $\forall t \geq 0$, where $\bar{\mathcal{U}}_2$ is a user-defined positive constant.


\section{Proposed Methodology}


Consider the LTI plant given in (\ref{plant}) and the reference model given in (\ref{ref}). Let, $\Omega^{'}_{U_1}:\{u\in\mathbb{R}^{m}: u^TMu < \bar{\mathcal{U}}_1^{'^2}\}$ and $\Omega^{'}_{U_2}:\{\dot{u}\in\mathbb{R}^{m}: \dot{u}^TM\dot{u} < \bar{\mathcal{U}}_2^{'^2}\}$
be two open sets, where $\bar{\mathcal{U}}_1^{'}=\bar{\mathcal{U}}_1\sqrt{\lambda_{min}\{M\}}$, $\bar{\mathcal{U}}_2^{'}=\bar{\mathcal{U}}_2\sqrt{\lambda_{min}\{M\}}$ and $M\in \mathbb{R}^{m\times m}$ is a user-defined positive definite matrix. We design the control input $u(t)\triangleq[u_1(t),\hdots, u_m(t)]^T$ to follow the  dynamics 
\begin{align}
    &\ddot{u}+\dot{u}+\alpha u=K_uv && u(0)\in \Omega_{U_1}^{'}, \dot{u}(0)\in\Omega_{U_2}^{'}\label{udot}
\end{align}
where the input rate $\dot{u}(t) \triangleq [\dot{u}_1(t), \hdots, \dot{u}_m(t)]^T \in \mathbb{R}^m$, $K_u(t)\in\mathbb{R}^{m\times m}$ is a subsequently designed controller parameter, $\alpha=\frac{\bar{\mathcal{U}}_2^{'^2}-\dot{u}^TM\dot{u}}{\bar{\mathcal{U}}_1^{'^2}-u^TMu}$. $v(t)\triangleq[v_1(t),\hdots, v_m(t)]^T \in \mathbb{R}^m$ is an auxiliary control input given by
\begin{align}
&v=\hat{K}_xx+K_rr
\label{veq1}
\end{align}
where $K_r\in \mathbb{R}^{m\times m}$ is a controller parameter and  $\hat{K}_x(t)\in \mathbb{R}^{m\times n}$ is the estimate of the true controller parameter $K_x\in \mathbb{R}^{m\times n}$.
\begin{assumption}
There exist true controller parameters $K_x$ and $K_r$ that satisfy the following matching conditions.
\begin{align}
    &A+BK_x=A_r 
    \label{mc1}\\
    &BK_r=B_r
    \label{mc2}
\end{align}
and $\|K_x\|\leq\bar{K}_x$, $\|K_r\|\leq\bar{K}_r$ where $\bar{K}_x,\bar{K}_r>0$ are assumed to be known.
\end{assumption}
\begin{remark}
   It is standard practice in projection-based adaptive control literature \cite{lavretsky2012robust} to consider known upper bounds on controller parameters. In many practical applications, while the exact values of controller parameters may be unknown, a priori knowledge of the upper bounds of \( K_x \), is often available which is useful for designing adaptive update laws.
\end{remark}
The trajectory tracking error is defined as $e(t) \triangleq x(t)-x_r(t)$. Employing (\ref{veq1}), (\ref{mc1}) and (\ref{mc2}) the closed-loop error dynamics for (\ref{plant}) and (\ref{ref}) can be obtained as
\begin{align}
    \dot{e}=A_re+B\tilde{K}_xx+B\Delta u+d
    \label{edot}
\end{align}
Here, \(\tilde{K}_x(t) \triangleq \hat{K}_x(t) - K_x \in \mathbb{R}^{m \times n}\) represents the error in parameter estimation, while \(\Delta u(t) \triangleq u(t) - v(t)\) denotes the difference between the actual and auxiliary control inputs. The term `$B\Delta u(t)$' can be considered as a disturbance and to alleviate its impact the following dynamics of an auxiliary error signal $e_1(t)$ is considered.
\begin{align}
\dot{e}_1=A_re_1+B\Delta u && e_1(0)=0
    \label{e1dot}
\end{align}
The difference error $e_d(t)$, which is the difference between the actual error and the auxiliary error, is defined as $e_d(t) \triangleq e(t)-e_1(t)$, and the closed-loop difference error dynamics is given by 
\begin{align}
\dot{e}_d=A_re_d+B\tilde{K}_xx+d && e_d(0)=e(0)
  \label{eddot}
\end{align}

\begin{assumption}
There exists known positive constant $\bar{\mathcal{X}}_r\in \mathbb{R}$ such that the reference trajectory is bounded as
\begin{align}
    & \|x_r(t)\| \leq \bar{\mathcal{X}}_r < \bar{\mathcal{X}}
    \end{align}
\end{assumption}
\begin{remark}
    Assumption 2 is a standard assumption in BLF-based literature \cite{BLF}, ensuring that the reference model trajectory stays within user-defined limits, preventing state constraint violations.
\end{remark}
Provided Assumption 2 holds, the plant state constraint can be reformulated as a bound on the tracking error: \(\|e(t)\| < \bar{\mathcal{E}}\), \(\forall t \geq 0\), where \(\bar{\mathcal{E}} = \bar{\mathcal{X}} - \bar{\mathcal{X}}_r\) is a positive constant, known a-priori. Thus, \(\|e(t)\| < \bar{\mathcal{E}}\) implies \(\|x(t)\| < \bar{\mathcal{X}}\) for all \(t \geq 0\).
To ensure constraint satisfaction on $e_d(t)$, $u(t)$ and $\dot{u}(t)$ consider the following BLFs $V_1(e_d)$, $V_2(u)$ and $V_3(\dot{u})$ defined on the sets $\Omega^{'}_{e_d}:\{e_d\in\mathbb{R}^{n}: e_d^TPe_d < \bar{\mathcal{E}}_d^{'^2}\}$, $\Omega^{'}_{U_1}:\{u\in\mathbb{R}^{m}: u^TMu < \bar{\mathcal{U}}_1^{'^2}\}$ and $\Omega^{'}_{U_2}:\{\dot{u}\in\mathbb{R}^{m}: \dot{u}^TM\dot{u} < \bar{\mathcal{U}}_2^{'^2}\}$, respectively.
\begin{align}
    &V_1(e_d) \triangleq \frac{1}{2} \log{ \frac{\bar{\mathcal{E}}_d^{'^2}}{\bar{\mathcal{E}}_d^{'^2}-e_d^TPe_d}} \\
    &V_2(u) \triangleq \frac{1}{2} \log{ \frac{\bar{\mathcal{U}}_1^{'2}}{\bar{\mathcal{U}}_1^{'2}-u^TMu}}\\
    &V_3(\dot{u}) \triangleq \frac{1}{2} \log{ \frac{\bar{\mathcal{U}}_2^{'2}}{\bar{\mathcal{U}}_2^{'2}-\dot{u}^TM\dot{u}}}
\end{align}
If $e_d^TPe_d\rightarrow \bar{\mathcal{E}}_d^{'^2}$ or $u^TMu\rightarrow \bar{\mathcal{U}}_1^{'2}$ or $\dot{u}^TM\dot{u} \rightarrow \bar{\mathcal{U}}_2^{'^2}$ i.e.
when the constrained states $e_d(t)$ or $u(t)$ or $\dot{u}(t)$ approach the boundary of the respective safe sets, the corresponding BLF $V_1(e_d)\rightarrow \infty$, or $V_2(u)\rightarrow \infty$, or $V_3(\dot{u})\rightarrow \infty$, respectively.\\
The adaptive update laws are designed as 
\begin{subequations}
\begin{equation}
    \dot{\hat{K}}_x=\text{proj}_{\Omega_{1}}\bigg(-\frac{\Gamma_xB^TPe_dx^T}{\bar{\mathcal{E}}_d^{'^2}-e_d^TPe_d}-\sigma_x\Gamma_x\hat{K}_x\bigg)
    \label{proposedlaw_kx}
    \end{equation}
    \begin{equation}
        \dot{K}_u=-\frac{\Gamma_uM\dot{u}v^T}{\bar{\mathcal{U}}_2^{'2}-\dot{u}^TM\dot{u}}
        \label{proposedlaw_ku}
    \end{equation}
    \label{proposedlaw}
\end{subequations}
where  $\Gamma_x\in \mathbb{R}^{m \times m}$ is a positive-definite matrix and $\sigma_x$ is a positive constant. The projection operator \(\text{proj}_{*}(\cdot)\) \cite{lavretsky2012robust} ensures that the parameter update law \((\cdot)\) constrains the parameters within a convex and compact region, denoted by `\(*\)', in the parameter space. In this context, the convex function associated with the projection operator is chosen as \(f(\hat{K}_x) = \|\hat{K}_x\|^2\) and the convex and compact region is defined by $\Omega_1=\{\hat{K}_x\in\mathbb{R}^{m\times n}|\|\hat{K}_x\|^2 \leq \bar{K}_x^2\}$, which aligns with Assumption 1.

\begin{theorem}
Consider the dynamics of the MIMO LTI plant (\ref{plant}) and user-defined reference model (\ref{ref}). Provided Assumptions 1-2 and the following feasibility conditions C1-C2 hold, \\
\textbf{C1}: The true controller parameter $K_x\in \mathbb{R}^{m\times n}$ satisfies the following inequality.
\begin{align}
  \bar{K}_x<\frac{\rho}{\|B\|}
  \label{C11}
\end{align}
where \(\rho<|\max(\lambda_{\Re}\{A_r\})|\) is a positive constant.\\
\textbf{C2}: The constraint on the plant state $\bar{\mathcal{X}}\in \mathbb{R}$ satisfies
\begin{align}
   \bar{\mathcal{X}}>\frac{1}{\gamma}\bigg(\kappa+\frac{\|B\|}{\rho}\|u(t)\|+\frac{2\lambda_{max}\{P\}\bar{d}}{\lambda_{min}\{Q\}}\bigg)+ \bar{\mathcal{X}}_r
   \label{C21}
\end{align}
where $\kappa=\frac{\|B\|}{\rho}(\bar{K}_x\bar{\mathcal{X}}_r+\bar{K}_r\bar{r})$ and $\gamma=(1-\frac{\|B\|\bar{K}_x}{\rho})$ are positive constants,\\
the proposed controller (\ref{udot}), (\ref{veq1}) and the adaptive laws (\ref{proposedlaw}) ensure the following. 
\begin{enumerate}
    \item [(i)] The plant states stay within the user-defined safe set given by $\Omega_x:=\{x\in \mathbb{R}^n:\|x(t)\|< \bar{\mathcal{X}}\}.$
    \item [(ii)] The input magnitude stays within the user-defined safe set given by $\Omega_{U_1}:=\{u\in \mathbb{R}^m:\|u(t)\|< \bar{\mathcal{U}}_1\}.$
    \item [(iii)] The input rate stays within the user-defined safe set given by $\Omega_{U_2}:=\{\dot{u}\in \mathbb{R}^m:\|\dot{u}(t)\|< \bar{\mathcal{U}}_2\}.$ 
    \item [(iv)] All closed-loop signals remain bounded.
\end{enumerate}
\end{theorem}

\begin{proof}
The proof of Theorem 1 consists of three parts where we consider two separate Lyapunov analyses. In the first segment, the constraint on the tracking error is reformulated as a constraint on the difference error to handle input saturation, yielding two feasibility conditions. The second part employs two BLFs to enforce constraints on $\|u(t)\|$ and $\|\dot{u}(t)\|$, which proves (ii) and (iii) and this result is exploited in the final segment, where another BLF is considered to constrain the difference error $\|e_d(t)\|$ which subsequently proves (i).\\ 
To deal with the input saturation term `$\Delta u$' present in the closed loop error dynamics (\ref{edot}), we transform the constraint on $e(t)$ to $e_d(t)$, which requires the bound on $e_1(t)$. Solving the differential equation (\ref{e1dot}),
\begin{align}
\|e_1(t)\|\leq &\|\exp(A_rt)\|\|e_1(0)\|+\|B\|\|\Delta u(t)\|  \nonumber\\
&\bigg(\|\int_0^t\exp(A_r(t-\tau))d\tau \|\bigg)
\end{align}
Since $e_1(0)=0$ and $A_r$ is Hurwitz, employing the properties of matrix exponential yields
\begin{align}
    \|e_1(t)\|\leq \frac{\|B\|\|\Delta u(t)\|}{\rho}
    \label{e1eq1}
\end{align}
where $\rho<|\max(\lambda_{\Re}\{A_r\})|$. The bound on  $\Delta u$ can be obtained as 
\begin{align}
    \|\Delta u(t)\| 
     & \leq \|u(t)\|+\bar{K}_x(\|e(t)\|+\bar{\mathcal{X}}_r)+\bar{K}_r\bar{r}
     \label{deltau}
\end{align}
Substituting $(\ref{deltau})$ in $(\ref{e1eq1})$,
\begin{align}
    \|e_1(t)\|\leq \kappa + \frac{\|B\|}{\rho}(\|u(t)\|+\bar{K}_x\|e(t)\|)
    \label{e1eq2}
\end{align}
where $\kappa=\frac{\|B\|}{\rho}(\bar{K}_x\bar{\mathcal{X}}_r+\bar{K}_r\bar{r})$. 
Employing (\ref{e1eq2}) and the fact that $e(t)=e_d(t)+e_1(t)$, the constraint on $e_d(t)$ can be obtained as
\begin{align}
    \|e_d(t)\|< \gamma\bar{\mathcal{E}}-(\kappa+\frac{\|B\|}{\rho}\|u(t)\|)=\bar{\mathcal{E}}_d
    \label{eqqqq}
\end{align}
The conditions $\gamma=(1-\frac{\|B\|\bar{K}_x}{\rho})>0$ and 
$\bar{\mathcal{E}}_d>0$ lead to the feasibility conditions C1 and C2, respectively. Now, to determine the lower bound on the state constraint from \((\ref{eqqqq})\), we need to prove $\|u(t)\|<\bar{\mathcal{U}}_1$.
To ensure constraint satisfaction on the magnitude and rate of the control input, consider the candidate Lyapunov function $V_{\theta}(u,\dot{u}, K_u):\Omega_{U_1}^{'} \times \Omega_{U_2}^{'} \times 
\mathbb{R}^M \rightarrow \mathbb{R}_{+}$,
\begin{align}
    V_{\theta}(u,\dot{u},K_u)=&\frac{1}{2} \bigg [ \log{ \frac{\bar{\mathcal{U}}_1^{'2}}{\bar{\mathcal{U}}_1^{'2}-u^TMu}}+ \log{ \frac{\bar{\mathcal{U}}_2^{'2}}{\bar{\mathcal{U}}_2^{'2}-\dot{u}^TM\dot{u}}}\nonumber \\
    &+tr(K_u^T\Gamma_u^{-1}K_u) \bigg]
    \label{lyap1}
\end{align}
Taking the time-derivative of $V_{\theta}$ along the system trajectory,
\begin{align}
    \dot{V}_{\theta}=&\frac{u^TM\dot{u}+\dot{u}^TMu}{2(\bar{\mathcal{U}}_1^{'2}-u^TMu)} + \frac{\ddot{u}^TM\dot{u}+\dot{u}^TM\ddot{u}}{2(\bar{\mathcal{U}}_2^{'2}-\dot{u}^TM\dot{u})}
\nonumber\\
&+tr(K_u^T\Gamma_u^{-1}\dot{K}_u)
\label{vdot21}
\end{align}
Substituting (\ref{udot}) in (\ref{vdot21}),
\begin{align}
    \dot{V}_{\theta}=&\frac{u^TM\dot{u}+\dot{u}^TMu}{2(\bar{\mathcal{U}}_1^{'2}-u^TMu)} + \frac{1}{2(\bar{\mathcal{U}}_2^{'2}-\dot{u}^TM\dot{u})} \bigg[(K_{u}v \nonumber \\
    &-\alpha u-\dot{u})^TM\dot{u}+\dot{u}^TM(K_{u}v-\alpha u-\dot{u})\bigg]\nonumber\\
    &+tr(K_u^T\Gamma_u^{-1}\dot{K}_u)
\end{align}
Substituting $\alpha=\frac{\bar{\mathcal{U}}_2^{'^2}-\dot{u}^TM\dot{u}}{\bar{\mathcal{U}}_1^{'^2}-u^TMu}$ and the adaptive update law from (\ref{proposedlaw_ku}),
\begin{align}
   \dot{V}_{\theta}= -\frac{\dot{u}^TM\dot{u}}{\bar{\mathcal{U}}_2^{'2}-\dot{u}^TM\dot{u}} \leq 0
   \label{lyapfunc1}
\end{align}
Since $V_{\theta}$ in (\ref{lyap1}) is positive definite and $\dot{V}_{\theta}$ is a negative semi-definite function from (\ref{lyapfunc1}),  $V_{\theta}(t)\leq V_{\theta}(0)$ $\forall t\geq0$. As the BLF $V_{\theta}(u,\dot{u}, K_u)$ is defined in the region $\Omega_{U_1^{'}, U_2^{'},K_u}:= \{[u^T, \dot{u}^T , K_u^T] \in \Omega_{U_1}^{'} \times \Omega_{U_2}^{'} \times \mathbb{R}^M : u^TMu< \bar{\mathcal{U}}_1^{'2} , \dot{u}^TM\dot{u}< \bar{\mathcal{U}}_2^{'2} \}$, it can be inferred from \cite[Lemma~1]{BLF} that $u^TMu< \bar{\mathcal{U}}_1^{'2}$ and $\dot{u}^TM\dot{u}< \bar{\mathcal{U}}_2^{'2}$ which imply $u^TMu < \lambda_{min}\{M\} \bar{\mathcal{U}}_1^{2}$ and $\dot{u}^TM\dot{u} < \lambda_{min}\{M\} \bar{\mathcal{U}}_2^{2}$, respectively.
For any $M>0$, we can obtain $u^TMu \geq \lambda_{min}\{M\}\|u\|^2$ and $\dot{u}^TM\dot{u} \geq \lambda_{min}\{M\}\|\dot{u}\|^2$. Provided $u(0)\in \Omega_{U_1}^{'}$ and $\dot{u}(0)\in \Omega_{U_2}^{'}$, it can be proved that 
\begin{align}
    \|u(t)\|<\bar{\mathcal{U}}_1 && \|\dot{u}(t)\|<\bar{\mathcal{U}}_2 && \forall t \geq 0 
\end{align}
Substituting the bound of $\|u(t)\|$ in (\ref{eqqqq}), 
we choose $\bar{\mathcal{E}}_d$ such that it satisfies 
\begin{equation}
\bar{\mathcal{E}}_d=\gamma\bar{\mathcal{E}}-(\kappa+\frac{\|B\|}{\rho}\bar{\mathcal{U}}_1)
\label{edbar}
\end{equation} 
Now to bound the plant states within user-defined constraint,  
consider the candidate Lyapunov function $V_{\phi}(e_d,\tilde{K}_x):\Omega_{e_d}^{'}\times \mathbb{R}^N\rightarrow \mathbb{R}_{+}$ as,
\begin{align}
    V_{\phi}(e_d,\tilde{K}_x)=&
    \frac{1}{2} \bigg [ \log{ \frac{\bar{\mathcal{E}}_d^{'^2}}{\bar{\mathcal{E}}_d^{'^2}-e_d^TPe_d}}  
+tr(\tilde{K}_x^T\Gamma_x^{-1}\tilde{K}_x)\bigg]
    \label{lyap}
\end{align}
Taking the time-derivative of $V_{\phi}$ along the system trajectory and employing the adaptive update law \((\ref{proposedlaw_kx})\), 
\begin{align}
    \dot{V}_{\phi}\leq&-\frac{e_d^TQe_d}{2(\bar{\mathcal{E}}_d^{'^2}-e_d^TPe_d)}+\frac{e_d^TPd}{\bar{\mathcal{E}}_d^{'^2}-e_d^TPe_d}\nonumber \\
    &-\sigma_x tr(\tilde{K}_x^T\tilde{K}_x)-\sigma_x tr(\tilde{K}_x^T{K}_x)
    \\
    \leq &-\frac{\lambda_{min}\{Q\}\|e_d\|}{2(\bar{\mathcal{E}}_d^{'^2}-e_d^TPe_d)}\bigg(\|e_d\|-\frac{2\lambda_{max}\{P\}\bar{d}}{\lambda_{min}\{Q\}}\bigg)\nonumber \\
    &-\frac{\sigma_x}{2} \|\tilde{K}_x\|^2 +\frac{\sigma_x}{2}\|K_x\|^2
    \label{lyapfunc}
\end{align}
To prove a bounded result from (\ref{lyapfunc}) the following condition should hold.
\begin{align}
\bar{\mathcal{E}}_d>\frac{2\lambda_{max}\{P\}\bar{d}}{\lambda_{min}\{Q\}}
\label{edbar1}
\end{align}
From (\ref{edbar}) and (\ref{edbar1}), we obtain the feasibility condition C2. At $t=0$ (\ref{lyapfunc}) can be written as
\begin{align}
    \dot{V}_{\phi}(0)< -\alpha V_{\phi}(0)+c
    \label{uub}
\end{align}
where $\alpha=\min(\lambda_{min}\{Q\},{\sigma_x})$ and $c=\frac{\sigma_x}{2}\|K_x\|^2$ are positive constants. Using \cite[Lemma~1]{uub_blf}, it can be proved from (\ref{uub}) that $e_d(0^+)\in \Omega_{e_d}$ at $t=0^+$, provided $e_d(0)\in \Omega_{e_d}$ and (\ref{edbar1}) is satisfied. Repeating the above argument for all time instants,
\begin{align}
    \dot{V}_{\phi}(t)< -\alpha V_{\phi}(t)+c && \forall t \geq 0
    \label{uub1}
\end{align}
Integrating over $[0, t]$ the differential inequality in (\ref{uub1}) leads to the following exponentially convergent bound on $V_{\phi}(t)$.
\begin{align}
    V_{\phi}(t)<V_{\phi}(0)+ \frac{c}{\alpha} && \forall t \geq 0
    \label{uuball}
\end{align}
From (\ref{lyap}) and (\ref{uuball}), it can be shown that
\begin{align}
    \frac{1}{2} \log{ \frac{\bar{\mathcal{E}}_d^{'^2}}{\bar{\mathcal{E}}_d^{'^2}-e_d^TPe_d}}< V(0)+\frac{c}{\alpha}
    \label{eq1}
\end{align}
Taking exponential on both sides of (\ref{eq1}),
\begin{align}
    e_d^TPe_d 
    &< \lambda_{min}\{P\}\bar{\mathcal{E}}_d^{'^2} \sqrt{1-e^{-2 (V(0)+\frac{c}{\alpha})}}
    \label{ebound}
\end{align}
Now for any $P>0$, $e_d^TPe_d \geq \lambda_{min}\{P\}\|e_d\|^2$. Since $V_{\phi}(0), \alpha, c$ are positive constants, from (\ref{ebound}), it can be proved that 
$\|e_d(t)\|<\bar{\mathcal{E}}_d $ $\forall t \geq 0 $.
Further,
\begin{align}
    \|e(t)\|
    < & \bar{\mathcal{E}}_d + \kappa+\frac{\|B\|}{\rho}(\bar{\mathcal{U}}_1+ \bar{K}_x\|e(t)\|)
    \label{eboun}
\end{align}
Substituting the value of $\bar{\mathcal{E}}_d$ from (\ref{edbar}) in (\ref{eboun}), 
\begin{align}
   \|e(t)\|<\gamma \bar{\mathcal{E}}+\frac{\|B\|}{\rho}\bar{K}_x\|e(t)\| 
   \label{ebound11}
\end{align}
Since $\gamma=(1-\frac{\|B\|\bar{K}_x}{\rho})$, from(\ref{ebound11}) it can be proved that $ \|e(t)\|<\bar{\mathcal{E}}$ $\forall t\geq 0$,
i.e. the trajectory tracking error remains within the pre-specified bound: $e(t) \in \Omega_e$ $\forall t\geq0$. Further, since $x(t)=e(t)+x_r(t)$, $\|x_r(t)\|\leq \bar{\mathcal{X}}_r$ and $\|e(t)\|<\bar{\mathcal{E}}$, the proposed control policy ensures that the plant states remain confined within the user-defined safe limit, i.e. 
 \begin{align}
     \|x(t)\|<\bar{\mathcal{E}}+\bar{\mathcal{X}}_r< \bar{\mathcal{X}} && \forall t \geq 0
 \end{align}
 which implies satisfaction of state constraint, i.e. $\|x(t)\|<\bar{\mathcal{X}}$  $\forall t\geq 0$.\\
Since \(V_{\theta}(t), V_{\phi}(t) > 0\), \(\dot{V}_{\theta}(t)\leq 0\) and $ \dot{V}_{\phi}(t) \leq -\alpha V_{\phi}+c$ it follows that \(e_d(t)\), \(u(t)\), \(\dot{u}(t)\), \(\tilde{K}_x(t)\), and \(K_u(t)\) are all in \(\mathcal{L}_{\infty}\). Given that \(K_x\) is constant, \(\hat{K}_x(t)\) is also in \(\mathcal{L}_{\infty}\), ensuring \(v(t) \in \mathcal{L}_{\infty}\). From (\ref{eboun}), we conclude \(e(t)\) and \(x(t)\) belong to \(\mathcal{L}_{\infty}\), guaranteeing boundedness of all closed-loop signals. Additionally, from (\ref{lyapfunc1}) and (\ref{udot}), we find \(\dot{u}(t)\in\mathcal{L}_2\) and \(\ddot{u}(t)\in\mathcal{L}_{\infty}\), respectively, implying uniform continuity of \(\dot{u}(t)\). Subsequently, by employing Barbalat's Lemma \cite{slotine}, we can prove that \(\dot{u}(t)\) converge to zero as \(t \to \infty\).
\end{proof}
    
    \begin{remark}
    The feasibility condition C1 imposes an upper bound on $\bar{K}_x$ to regulate the saturation deficiency `$\|\Delta u(t)\|$', ensuring stabilization. A larger $\bar{K}_x$ leads to a smaller $\gamma$, thereby reducing the feasibility region. From (\ref{mc1}) and (\ref{C11}), we obtain $\|A_r-A\|\leq\|B\|\bar{K}_x<\rho$, implying that the eigenvalues of \( A \) must remain sufficiently close to those of \( A_r \), which is stable by design. While stability is necessary for feasibility, it is not always sufficient. The other feasibility condition in \( C2 \) ensures that the state constraint \( \bar{\mathcal{X}}\) is large enough to account for the effects of the input constraints and disturbances. Since \( \gamma < 1 \), the feasibility region of state constraint \( \bar{\mathcal{X}} \) may be conservative; however, this trade-off guarantees feasibility for any given input constraint and disturbance bound. Increasing the input constraint expands the region of feasible state constraint and the set of admissible initial conditions.  
    \end{remark}
\section{Simulation Results}
\begin{figure}[h!]
     \centering
     \includegraphics[width=0.85\linewidth]{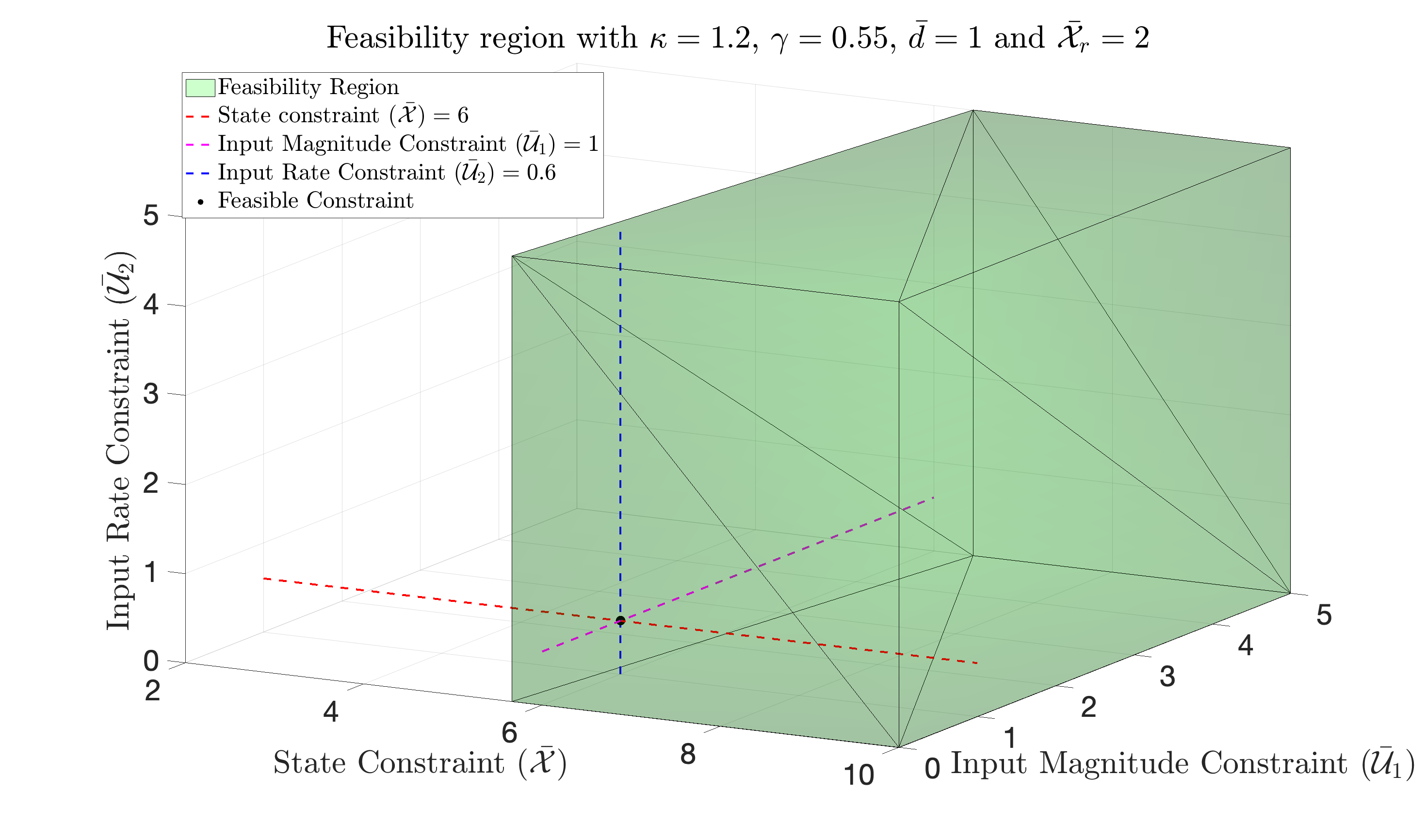}
     \caption{Feasibility Region}
     \label{feasibility_plot}
 \end{figure}

To validate the effectiveness of the proposed algorithm, we consider the lateral dynamics of an aircraft which follows (\ref{plant}), where $x(t)=[\theta(t),q(t),h(t),w(t)]^T\in \mathbb{R}^4$, $\theta(t)$ denotes the pitch angle, $q(t)$ represents the pitch rate, $h(t)$ denotes the altitude and $w(t)$ is the vertical velocity. The control input $u(t)=[\delta_e(t),\delta_t(t)]\in\mathbb{R}^2$ consists of elevator deflection ($\delta_e$) and throttle input ($\delta_t$), which influences the pitch and vertical motion, $d(t)\in \mathbb{R}^4$ represents the bounded external disturbances such as wind gusts, atmospheric disturbances etc.  
The desired stable reference model follows (\ref{ref}). The plant and reference model matrices are given as 
\begin{equation*}
\scriptstyle 
\begingroup 
\setlength\arraycolsep{1.5pt}
A=\begin{pmatrix}
    \scriptstyle  0 & \scriptstyle 4 & \scriptstyle  0 & \scriptstyle 0 \\
\scriptstyle -15 & \scriptstyle -15.85& \scriptstyle  -4.02 &\scriptstyle  -5.7 \\
   \scriptstyle 0 &\scriptstyle 0 &\scriptstyle 0 &\scriptstyle 4 \\
   \scriptstyle -6.85 &\scriptstyle  -9.9 &\scriptstyle -8 &\scriptstyle  -9.8 
    \end{pmatrix} \endgroup \hspace{6pt}
    \begingroup 
\setlength\arraycolsep{1pt}    
    \scriptstyle B=\begin{pmatrix}
    \scriptstyle 0 & \scriptstyle 0\\
    \scriptstyle 0.2 & \scriptstyle 0\\
    \scriptstyle 0 & \scriptstyle 0\\
    \scriptstyle 0 & \scriptstyle 0.2\\ 
\end{pmatrix} 
\endgroup
\end{equation*}

\begin{equation*}
\scriptstyle 
\begingroup 
\setlength\arraycolsep{1.5pt}
A_r=\begin{pmatrix}
    \scriptstyle  0 & \scriptstyle 4 & \scriptstyle  0 & \scriptstyle 0 \\
\scriptstyle -14.18 & \scriptstyle -16.05& \scriptstyle  -3.88&\scriptstyle  -6.12\\
   \scriptstyle 0 &\scriptstyle 0 &\scriptstyle 0 &\scriptstyle 4 \\
   \scriptstyle -7 &\scriptstyle  -10.2 &\scriptstyle -7 &\scriptstyle  -10.2 
    \end{pmatrix} \endgroup \hspace{4pt}
    \begingroup 
\setlength\arraycolsep{1pt}    
    \scriptstyle B_r=\begin{pmatrix}
    \scriptstyle 0 & \scriptstyle 0\\
    \scriptstyle 1 & \scriptstyle 0\\
    \scriptstyle 0 & \scriptstyle 0\\
    \scriptstyle 0 & \scriptstyle 2\\ 
\end{pmatrix} 
\endgroup
\end{equation*} 
For simulation, the parameters are chosen as: $r(t)=[0.4\sin(t/10);0.2\cos(t/20)]$, $\Gamma_x=5\mathbb{I}_{2\times 2}$, \(\sigma_x=1\), $\Gamma_u=2I_{4 \times 4}$, 
 $Q=\mathbb{I}_{4 \times 4}$, $M=\mathbb{I}_{2 \times 2}$ and $\bar{\mathcal{X}}_r<2$. The user-defined constraints on the input magnitude and the input rate are given as $\bar{\mathcal{U}}_1=1$ and $\bar{\mathcal{U}}_2=0.6$, respectively. The bound on the unknown external disturbance is given as $\bar{d}=1$. Since \(\rho < |\max(\lambda_{\Re}\{A_r\})| < 2.4\), we select \(\rho = 2.3\). Using the feasibility condition from equation \((\ref{C11})\), we determine that \(\bar{K}_x < 11.5\). One way to estimate \(\bar{K}_x\) is to use the knowledge of upper bound of \(\|B^{\dagger}(A_r - A)\|\), where \(B^{\dagger}\) is the left inverse of \(B\), given by \(B^{\dagger} = (B^T B)^{-1} B^T\). We verify that the feasibility condition \(C1\) is satisfied since \(\|B^{\dagger}(A_r - A)\| < 11.5\). For simulation purposes, we have chosen \(\bar{K}_x = 5\) and \(\bar{K}_r = 10\), with parameters \(\gamma = 0.55\) and \(\kappa = 1.2\). Applying C2, the lower bound for the state constraint is \(\bar{\mathcal{X}} > 5.8\). Fig. \ref{feasibility_plot} shows the feasibility region for the given plant and reference model. We can observe from Fig. \ref{feasibility_plot} that without a hard constraint on state, the controller remains feasible for any input magnitude and rate constraints. For simulation, we consider $\bar{\mathcal{X}}=6$. Transforming the state constraint, we set \(\bar{\mathcal{E}} = 4\) and \(\bar{\mathcal{E}}_d = 0.9\). 
 To demonstrate the efficacy of the proposed control law, we conduct a comparison with the robust MRAC approach \cite{classMRAC}. The adaptive gains for the robust MRAC are selected as \( \Gamma_x = 15 \mathbb{I}_2 \) and \(\sigma_x=1\). Note that, the adaptation gains for both the proposed controller and the robust MRAC are carefully tuned to ensure satisfactory tracking performance.

\begin{figure}[h!]
         \centering
         \includegraphics[width=0.85\textwidth]{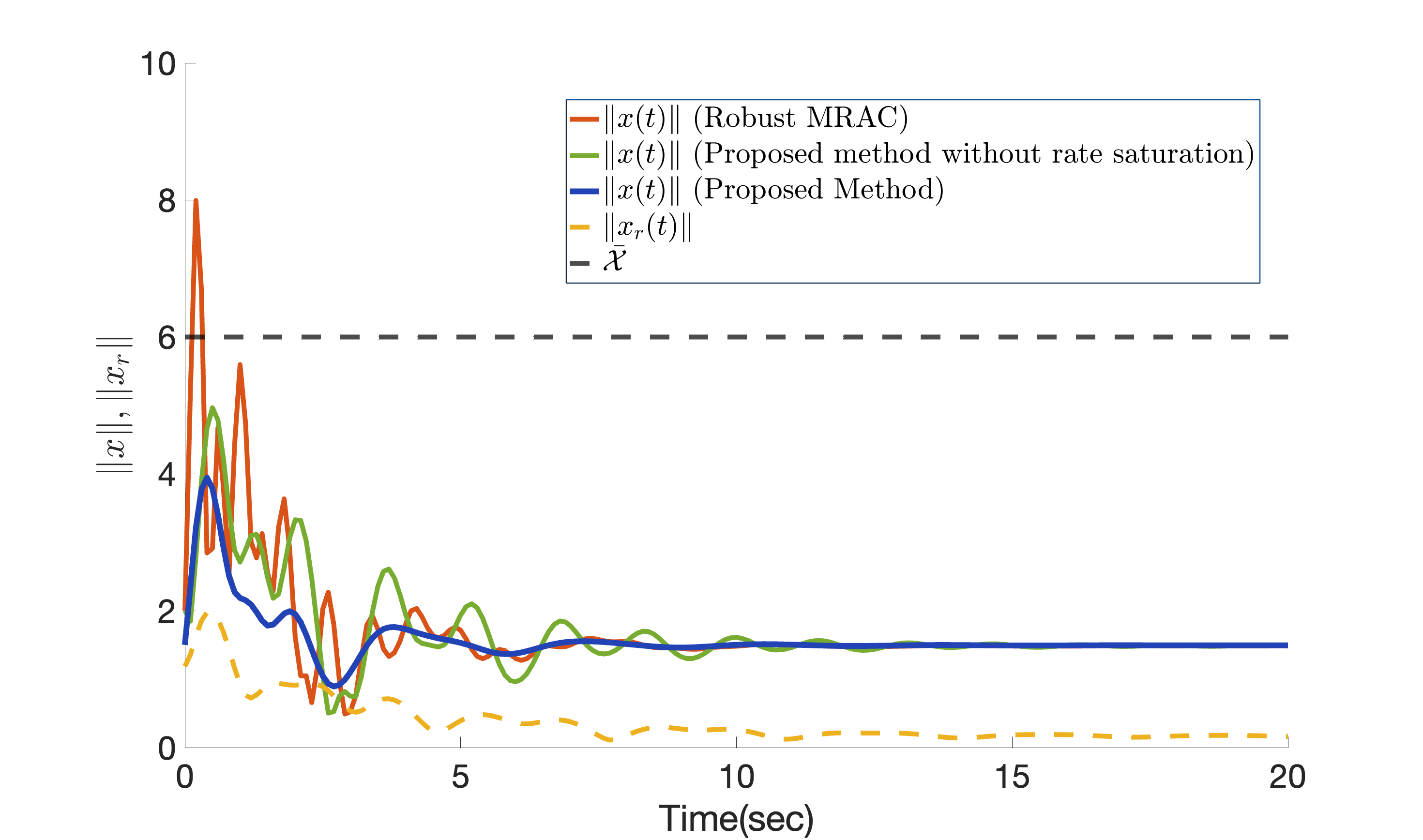}
     \caption{Comparison of tracking performance with the proposed controller and robust MRAC.}
     \label{tracking}
\end{figure}
\begin{figure}[h!]
    \centering
    \includegraphics[width=0.85\linewidth]{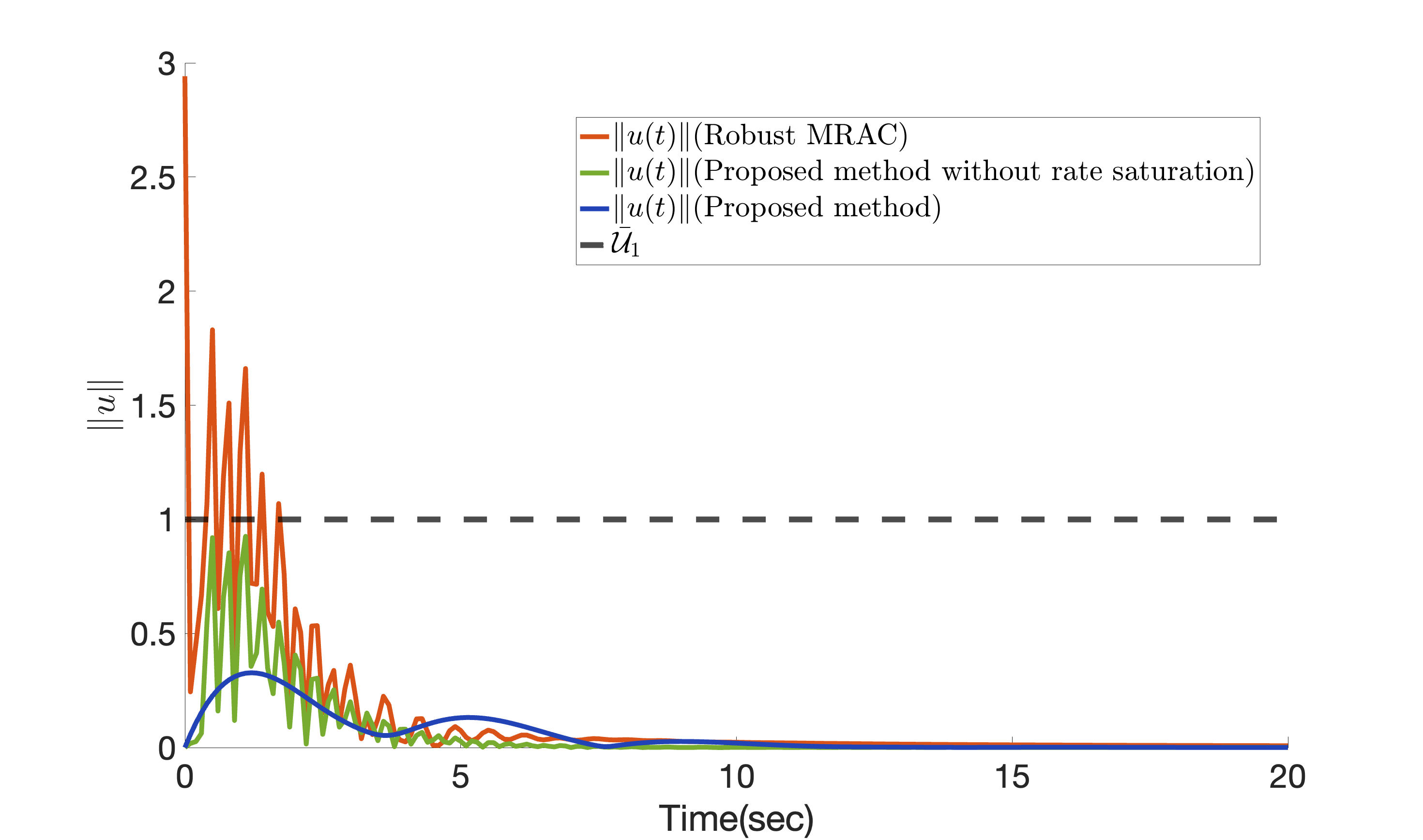}
    \caption{Comparison of input magnitude with the proposed controller and robust MRAC.}
    \label{u1a}
\end{figure}

\begin{figure}[h!]
    \centering
    \includegraphics[width=0.88\linewidth]{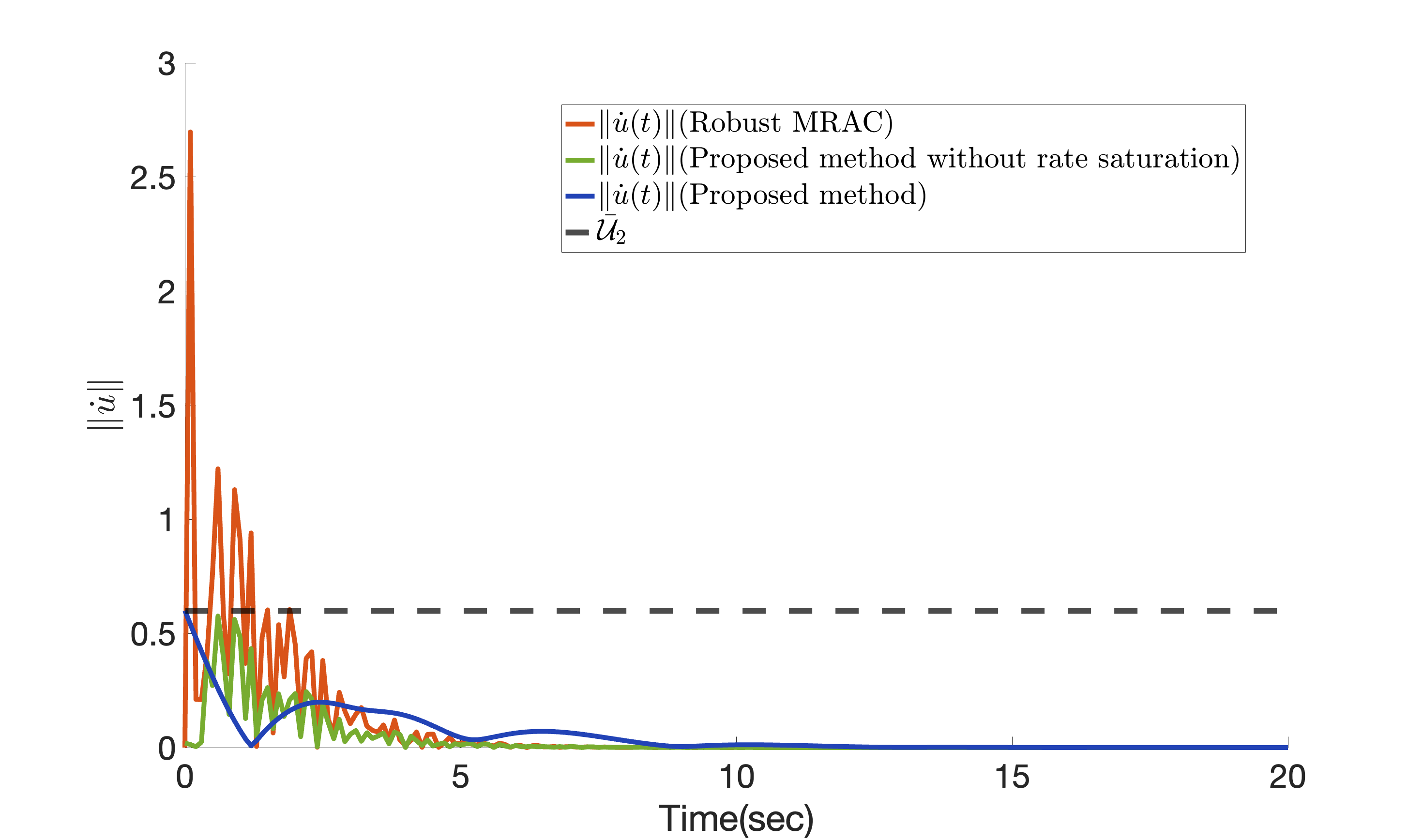}
    \caption{Comparison of input rate with the proposed controller and robust MRAC.}
    \label{u1b}
\end{figure}


Fig. \ref{tracking} shows that the proposed method guarantees state constraint satisfaction, i.e., \(\|x(t)\| < 6\), which implies \(\|e_d(t)\| < 0.9\) and \(\|e(t)\| < 4\) for all \(t \geq 0\). In contrast, in the case of robust MRAC, the plant state exceeds the user-defined safe set boundaries. Since \(\|e_1(0)\| = 0\), we have \(\|e_d(0)\| = \|e(0)\|\), ensuring constraint satisfaction with \(\|e(0)\| < 0.9\).

The proposed controller also limits control input magnitude (Fig. \ref{u1a}) and rate (Fig. \ref{u1b}) within specified bounds, i.e., \(\|u(t)\| < 1\) and \(\|\dot{u}(t)\| < 0.6\), while the robust MRAC violates these constraints. Without constraining the input rate for both the proposed method (with only state and input magnitude constraint) and robust MRAC, the magnitude of the control input, although bounded within the pre-specified region, exhibits significant oscillations, which can lead to undesirable system behavior and reduced performance. In contrast, imposing a rate constraint mitigates these oscillations, resulting in a smoother control input and system response, which enhances the performance and reliability of the control system.

\section{Conclusion}
In this paper, we present a novel robust MRAC architecture for uncertain MIMO LTI systems that considers state, input magnitude, and rate constraints in the presence of bounded disturbances. Our approach offers a practical alternative to optimization-based constrained control methods, which can be computationally intensive and overly conservative for uncertain plants. A major contribution of the work is to provide verifiable conditions to check the feasibility of the control policy. While the proposed design applies to stable systems, it is the first approach to achieve these guarantees without requiring an optimization routine. Extending this framework to handle unstable plants is a key direction for future research. We demonstrate the effectiveness of the proposed control law through a comparative simulation study with robust MRAC on an aircraft dynamics model.


\bibliographystyle{ieeetr}
\bibliography{ref}

\end{document}